\documentclass[aps, twocolumn,
showpacs]{revtex4}
\usepackage{mathrsfs}
\usepackage{pifont}
\usepackage{graphicx}
\usepackage{amsmath,amssymb,amsthm}
\usepackage[colorlinks=true,urlcolor=blue,linkcolor=blue,citecolor=red]{hyperref}
\newcommand{\beq}{\begin{equation}}
\newcommand{\eeq}{\end{equation}}
\newcommand{\bea}{\begin{eqnarray}}
\newcommand{\eea}{\end{eqnarray}}

\begin{document}

\draft
\title{Close to the edge: hierarchy in a double   braneworld}

\author{Rommel Guerrero $^1$, Alejandra Melfo $^2$, Nelson Pantoja $^2$
and R. Omar Rodriguez $^1 $}
\address{ $^1$ Unidad de Investigaci\'on en Ciencias Matem\'aticas,
Universidad Centroccidental Lisandro Alvarado, 400 Barquisimeto,
Venezuela}
\address{ $^2$ Centro de F\'isica Fundamental, Universidad de Los Andes,
M\'erida, Venezuela}

\begin{abstract}

We show that the hierarchy between the Planck and the weak scales
can follow from the tendency of gravitons and fermions to localize
at different edges of a thick double wall embedded in an $AdS_5$
spacetime without reflection symmetry. This double wall is a stable
BPS thick-wall solution with two sub-walls located at its edges;
fermions are coupled to the scalar field through Yukawa interactions, but 
 the lack of reflection symmetry forces them
to be localized in one of the sub-walls.
We show that the graviton zero-mode wavefunction is suppressed in the
fermion edge by an exponential function of the distance
between the sub-walls, and that the massive modes
 decouple so that Newtonian gravity is recuperated.

\end{abstract}


\maketitle

\vspace{0.3cm}

\section{Introduction}

The idea of confining our four-dimensional Universe inside a
topological defect embedded in a higher dimensional spacetime dates
at least as far back as the suggestion of  Rubakov and Shaposhnikov 
\cite{Rubakov:1983bb} (see also \cite{Akama:1982jy}) that we could be living inside a domain wall.
They showed that fermions (of one chirality) can be confined to the
wall by their Yukawa interactions with the scalar field. Domain
walls are vacuum configurations that are topologically protected and
thus stable, but their gravitational interactions cannot be ignored.
This became evident after the work of Randall and Sundrum
\cite{Randall:1999vf}, in which they showed that the
five-dimensional  metric  produced by an infinitely thin brane is
enough  to ensure that Newtonian gravity is reproduced on the brane.

Actually, the RS brane does not even have to be a true domain wall,
in the sense that no scalar field is needed for the effect. One just
has to ensure that the bulk spacetime is Anti-De Sitter (AdS), and
on the brane the four-dimensional cosmological constant is set to
zero. The natural question is then whether fermions can also benefit
from this effect, allowing us to live in a domain wall just because
of its self-gravitation. The  answer is on the negative, as shown in
\cite{Bajc:1999mh} fermion modes  behave exactly opposite as the
gravitons,  the warp factor forcing them to escape from the wall
into the bulk. If one wants matter to be confined to the wall, it is
necessary  to combine Rubakov-Shaposhnikov with Randall-Sundrum,
i.e., to consider a real domain wall, made up of the vacuum
expectation value of a scalar field that breaks a discrete symmetry,
take into account its gravitational self-interactions, and make it
couple to the fermions. This amounts to solve the five-dimensional
coupled Einstein-Klein-Gordon system for an adequate potential, and
many such solutions can be found in the literature. Among them, BPS
walls, those where the four-dimensional cosmological constant is set
to zero, are the most appealing ``thick brane" generalizations of the
Randall-Sundrum scenario. BPS solutions to the Einstein-Klein-Gordon
system can be found by means of an auxiliary function of the scalar
field, the fake superpotential of the first-order formalism of
\cite{Behrndt:1999kz,Skenderis:1999mm,DeWolfe:1999cp}.

The fact that fermion zero mode localization requires exactly the
inverse warp factor as graviton zero mode localization, is used in
this paper to provide a rationale for the large hierarchy between
the Planck and weak scales in some particular thick wall solutions.
These solutions are a straightforward generalization of the simplest
BPS wall which represents a smoothing of the scenario of
\cite{Randall:1999vf}, found by Gremm \cite{Gremm:1999pj}.  
The generalization produces an asymmetric double wall system: 
a   BPS wall with a substructure consisting of two sub-wall located at its edges, 
 but  with different bulk cosmological constants on both sides.   
 As a consequence of the asymmetry,
fermions coupled to the scalar field are forced by the warp factor
to localize on one of the sub-walls, the fermion sub-wall. On the
other hand, the graviton zero-mode wavefunction is suppressed in the
fermion sub-wall by an exponential function of the wall's thickness,
i.e. the distance between the sub-walls, and gravitons localize in
the opposite sub-wall, the Planck sub-wall. This allows one to
provide a large hierarchy between the effective Planck masses on
both sub-walls, as proposed by Randall and Sundrum in their earlier
work \cite{Randall:1999ee} but with no orbifold geometries and no
negative tension branes.

Attempts to achieve suppressed mass scales with two positive tension
branes, the so-called Lykken-Randall scenario \cite{Lykken:1999nb},
are found in the literature \cite{losotros}. To our knowledge,
however, all of them require some form of radion field in order to
stabilize the extra dimension \cite{Kanti:2002zr}. In our case, the
stability of the two wall system stems from their topological
properties, they are just a special kind of BPS walls. The fact that
no compactification of the bulk coordinate is required allows us to
reproduce Newtonian gravity on the walls as in
\cite{Randall:1999vf}, while keeping a large mass hierarchy as in
\cite{Randall:1999ee}. Moreover, the fermions are not arbitrarily
assumed to be located in a different wall as the gravitons, they do
so as a consequence of spacetime being warped. In contrast with the scenario of
 \cite{Lykken:1999nb}, the fermion  sub-wall  is not a  ``probe" brane,  and it has a
  non-negligible  tension.  

The paper is organized as follows. In the next Section we give an
overview of the mechanism that provides stable, asymmetric double
walls from a  known BPS solution,  and how fermions get  localized
on the   brane with larger Planck mass.  The following section is
dedicated to explicitly construct these solutions from the simplest
known BPS wall of \cite{Gremm:1999pj},  to find the graviton and fermion zero modes,
 and to calculate  the
Newtonian potential.

\section{Double walls and  hierarchy  }\label{set-up}

\begin{figure}
\includegraphics[width=6cm,angle=0]{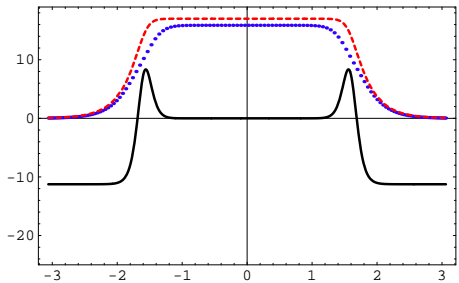}
 \caption{ Energy density (continuous), gravitational zero mode (dashed)
and fermion zero mode (dotted) for the double branes with $\alpha>0, \beta=0,\delta=4,s=11, \lambda=4\alpha $, and  arbitrary normalization.}
\label{fig1}
\end{figure}

The so-called BPS double walls are solutions to the 5-dimensional
Einstein-scalar field set of equations that satisfy the BPS
condition. As such, for the line element written in ``proper
length'' coordinates
 \beq
  \label{propermetric}
  ds^2 = e^{2 A(\xi)}\eta_{\mu\nu} dx^\mu dx^\nu + d\xi^2
 \eeq
they can be generated from a ``fake superpotential'' $W(\phi)$
\cite{Behrndt:1999kz,Skenderis:1999mm,DeWolfe:1999cp} by solving the
BPS equations for the scalar field, warp factor and scalar potential
\begin{eqnarray}
\phi^{\prime} &=& 3 \frac{d W}{d\phi}  \nonumber\\
A^{\prime} &=& -  W \nonumber \\
V(\phi)& =& \frac{3}{2}  \left[ 3 \left( \frac{d W}{d\phi} \right)^2
-  4 W^2 \right] \label{bps}
\end{eqnarray}
where prime denotes derivative with respect to the bulk coordinate
$\xi$. For $W(\phi)$ given by
 \beq\label{W}
 W (\phi)=\alpha[\sin(\phi/\phi_0)]^{2 - 1/s},
 \eeq
with $\alpha$ and $\phi_0$  real constants, solutions to (\ref{bps})
were found in \cite{Melfo:2002wd} representing a family of double
branes parametrized by an odd integer $s>1$ that interpolate between
$AdS_5$ spacetimes. For $s=1$, this is just the brane of Ref.
\cite{Gremm:1999pj}, i.e. a regularization of the infinitely-thin RS
brane, but for $s>1$ the wall splits in two in a well defined sense:
the energy density has two maxima as can be seen in Fig. \ref{fig1}.
The fact that exponentiating the superpotential for a single wall
gives rise to double systems was also used in \cite{CBazeia:2003qt}
to construct BPS double walls. The topological charge is given by
the asymptotic values of the superpotential
 \beq
 Q_T = W(+\infty) - W (-\infty) = 2\alpha
 \eeq
and is independent of $s$, i.e., it is the same as in the single
wall.

We have then a stable wall with two sub-walls, in
which the thickness of the sub-walls goes to zero as $s\rightarrow
\infty$, while the separation between them remains fixed. The
spacetime far away of the wall is $AdS_5$ with the same cosmological
constant ($\Lambda_\pm$) on both sides,  while the spacetime in
between the sub-branes is nearly flat, i.e.
 \beq
 \Lambda_\pm = - 6 \alpha^2, \qquad \Lambda_{\rm in} = 0
 \eeq

The gravitational zero modes are calculated as usual \beq \psi^g_0 =
N_g \, e^{ 2 A(\xi)} \eeq with $N_g$ a normalization factor. Because
the spacetime between the two sub-branes is nearly flat, the zero
modes are not peaked at $\xi=0$, but instead distribute smoothly
over the whole system \cite{Castillo-Felisola:2004eg}, as seen in
Fig. \ref{fig1}. A similar behavior has been found for other BPS
double branes \cite{CBazeia:2003qt}.

Fermion modes of a given chirality can also be localized, by adding
as usual a Yukawa coupling $\lambda$ with the scalar field
\cite{Bajc:1999mh}. One obtains
 \beq
 \psi_0^f(\xi) \sim \, e^{- 2 A(\xi) } e^{-\lambda \int \phi(\xi) d\xi}
 \eeq
The fermion zero modes for the system considered have been
calculated in \cite{Melfo:2006hh}, and can be seen in Fig.
\ref{fig1}. Details of the calculation are given in the next
section, but the general behavior will suffice for the time being.

Now, suppose the superpotential is shifted by a positive constant
 \beq
 \tilde W = W + \beta \label{Ws}
 \eeq
According to (\ref{bps}), we will have the same solution for the
scalar field. Furthermore, since for the double wall
(\ref{bps},\ref{W}), the extrema of $V(\phi)$ are the same ones of
$W(\phi)$, it follows that the extrema of the new scalar potential
$\tilde V(\phi)$ are the same ones of $V(\phi)$, namely $\phi=0$ and
$\phi= \pm\pi \phi_0/2$. Hence, the two sub-walls are situated at
the same place as before. Notice that the topological charge is the
same. However, the cosmological constants are now different at the
two sides of the double wall and the spacetime in between the
sub-walls is no longer flat, one has
 \beq\label{asimcosmo}
 \Lambda_-= -6(\alpha - \beta)^2,\quad\Lambda_+=
 -6(\alpha + \beta)^2,\quad\Lambda_{\rm in}= - 6 \beta^2
 \eeq
Accordingly, the new warp factor is from (\ref{bps})
 \beq
 \tilde A(\xi) = A(\xi) -\beta \xi
 \eeq
As long as we keep $\beta < \alpha$, the gravitational zero modes
remain localized, but now they are exponentially suppressed for $\xi
> 0$
 \beq
 \tilde{\psi}_0^g \sim \,\psi_0^g \, e^{-2 \beta \xi}
 \eeq
so that the graviton zero-mode function gets shifted towards the
region with a smaller cosmological constant, i.e. with $\Lambda_-$.

\begin{figure}
\includegraphics[width=6cm,angle=0]{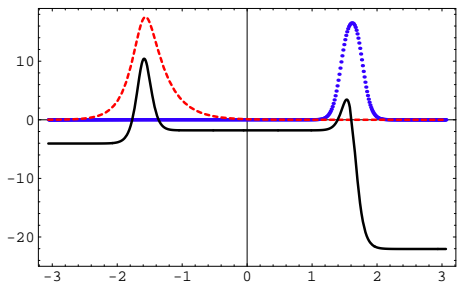}
\caption{ Energy density (continuous), gravitational zero mode (dashed)
and fermion zero mode (dotted) for the double branes with $\alpha>0, \beta=0.8\alpha,\delta=4,s=11, \lambda=4\alpha$, and  arbitrary normalization.}
\label{fig2}
\end{figure}

Fermion modes behave in exactly the opposite way, they are now
 \beq
 \tilde {\psi}^f_0 \sim \,\psi^f_0 \, e^{2 \beta \xi }
 \eeq
so that they are shifted towards the larger cosmological constant
region, i.e. with $\Lambda_+$. In other words, the gravitons are
localized on the sub-brane situated at $\xi_-$, henceforth the Planck
brane, while the fermion zero modes do the opposite, they live in
the opposite sub-brane around $\xi_+$, the weak brane. The situation
is depicted in Fig. \ref{fig2}.

The Planck masses in the Planck brane, $M_{Pl}^{\rm Planck} $, and in the weak
brane, $M_{Pl}^{\rm Weak}$, are approximately related by
 \beq
 M_{Pl}^{\rm Planck}   =   M_{Pl}^{\rm Weak} e^{-2 \beta \Delta}
 \label{supress}
 \eeq
with $\Delta$ the inter-sub-brane separation. Since $\Delta$ here is  a coordinate-dependent quantity,
 Eq.(\ref{supress})
should  be made precise by calculating and comparing the
gravitational potential on both sub-branes, which we do in the next
section.

As the thickness of the sub-branes approaches zero, this solution
amounts to having a thick domain wall with two infinitely-thin
sub-branes, the Planck brane and the weak brane, situated at
different edges (in the transverse direction) of the wall. Clearly,
the crucial role here is played by the double-brane solution, for
which BPS stability arguments apply \cite{Skenderis:1999mm}.

\section{Explicit solutions}

\subsection{Double branes}

Solutions to the system (\ref{bps}) with un-shifted superpotential
(\ref{W}) (i.e. reflection symmetry) have been found in
\cite{Melfo:2002wd} by switching to the so-called ``gauge''
coordinates, where the metric is written as
 \beq \label{gaugemetric}
  ds^2 = e^{2 A(y)} \eta_{\mu\nu} dx^\mu dx^\nu +  e^{2 H(y)} dy^2
 \eeq
The field equations can be integrated to give
\begin{eqnarray}\label{doublesol}
H(y)&=& - \frac{1}{2s}  \ln\left[1+\left(\frac{\alpha
y}{\delta}\right)^{2s}\right] , \qquad A(y) = \delta H(y)
\nonumber\\
V(\phi)&=&3\alpha^2[\sin(\phi/\phi_0)]^{2-2/s}\left[\frac{2s+4\delta-1}{2\delta}
\cos^2(\phi/\phi_0) - 2\right]
\nonumber\\
\phi(y)&=&\phi_0\arctan\left( \frac{\alpha y}{\delta}\right)^s,\quad
\phi_0= \frac{\sqrt{3\delta (2s -1)}}{s}
\end{eqnarray}

 The parameter $\delta$
gives the thickness of the wall. For $s=1$, this is just the brane
of Ref. \cite{Gremm:1999pj} and it can be rigorously shown that the
distributional limit $\delta \to 0$ gives the RS spacetime
\cite{Guerrero:2002ki}. But for non-zero $\delta$ and $s>1$, the
energy density has two maxima, as can be seen in Fig. \ref{fig1}.
Other double wall systems have been found in \cite{CBazeia:2003qt}
and \cite{Guerrero:2005aw}.

These walls have been studied in detail in a series of papers, their
thin wall limit $\delta\rightarrow 0$ in \cite{Melfo:2002wd},
localization of gravity in \cite{Castillo-Felisola:2004eg} and of
chiral fermion modes in \cite{Melfo:2006hh}. For $\delta=1$, there
is a normalizable zero mode and a continuous of massive states that
asymptote to plane waves as $y\rightarrow \pm \infty$. This
continuum of modes decouple, in the sense that they generate small
corrections to the Newtonian potential of the wall, at scales larger
than the fundamental length-scale of the system, $\alpha^{-1}$.

Let us  consider the double domain wall
(\ref{gaugemetric}, \ref{doublesol}) for $s\gg1$. The separation
between the sub-branes, i.e. the distance between the maxima of the
energy density, is given by
 \beq
 \Delta=2\frac{\delta}{\alpha}\left(\frac{s-1}{s+2\delta}\right)^{1/{2s}}
 \eeq
which behaves as $2{\delta}/{\alpha}$ for $s\gg 1$, while their
thickness $\gamma$ is approximately given by
 \beq
\gamma \sim \frac{1}{\alpha} \frac{\delta}{s} \ln\left(\frac{s}{\delta}\right) + O(s^{-2})
 \eeq
from which follows that $\gamma \sim 0$ for $s\gg 1$. Note that the
double wall is lost in the $\delta\rightarrow 0$ limit
\cite{Melfo:2002wd}, as expected from the fact that the
distributional geometry $\delta\rightarrow 0$ may be identified with
the asymptotic behavior (i.e. far away from the wall) of the domain
wall spacetime \cite{Pantoja:2003zr}.

Following \cite{Guerrero:2002ki}, the distributional limit $s \to
\infty$ of the Einstein tensor is found to be
 \begin{eqnarray}
 \label{double-eins}
&&\lim_{s\rightarrow\infty}\left(G^a_{\,\,b} + \Lambda g^a_{\,\,b}\right)= \nonumber\\&&-3\alpha
 \left[ \delta(y-y_+) + \delta(y-y_-)\right]\left(\partial^adt_b +
 \partial^b_{x^i}dx^i_b\right)
 \end{eqnarray}
with $\Lambda$ given by
 \beq
- \frac{\Lambda}{6}=\begin{cases}
        \alpha^2,& |y|>\delta/\alpha \cr
        0,& |y|<\delta/\alpha \cr

        \end{cases}
 \eeq
Hence for $s\gg1$ we have two infinitely thin walls located at
$y_+=\delta/\alpha$ and $y_-=-\delta/\alpha$, with tensions
$\sigma_+=\sigma_-=3\alpha >0$. On the other hand, we find that the
limit ${s\rightarrow \infty}$ of the metric tensor gives the line
element
  \beq
 \label{inf-double}
   ds^2 =\left\{\begin{array}{lr}  \left(\frac{\delta}{\alpha y}\right)^{2\delta}(-dt^2+dx^idx^i) +
        \left(\frac{\delta}{\alpha y}\right)^{2}dy^2 &,  
        |y|>{\delta}/{\alpha}
        \\  &   \\
     -dt^2 +dx^idx^i + dy^2 &,  |y|<\delta/\alpha  \end{array}\right.
\eeq
 and the distributional Einstein tensor (\ref{double-eins}) turns out
to be the Einstein tensor of the limit metric (\ref{inf-double}).

\subsection{Asymmetric double branes}

Now, shifting the superpotential as in (\ref{Ws}) results in a
solution with \bea\label{adw}
\tilde H(y)&=& H(y) \nonumber \\
\tilde A(y) &=& A(y)-\beta y\mathcal{F}(y)\nonumber\\
\tilde V(\phi)&=& V(\phi) - 6 \beta [ \beta + 2 \alpha
(\sin(\phi/\phi_0))^{2-1/s} ]
\nonumber\\
\tilde \phi(y)&=& \phi(y) \eea where
$\mathcal{F}(y)\equiv\,_2F_1[1/2s,1/2s,1+1/2s,-(\alpha
y/\delta)^{2s}]$ is the hypergeometric function. For $|\alpha
y/\delta| \leq 1$, i.e. inside the double brane, and large $s$,
$_2F_1\,\sim 1$ is a very good approximation, although we have used
the exact solution whenever performing numerical integrations.

The equation for the gravitational zero modes is found as usual. In
the axial gauge $h_{\mu y}=0$, writing the transverse-traceless part
of the metric perturbations $h_{\mu\nu}$  as
 \beq
 h_{\mu\nu}   = e^{i p\cdot x}  e^{H/2 } \psi_{\mu\nu}
 \eeq
one obtains
 \beq
 \left( - \frac{d^2}{dy^2} + U(y) \right) \psi_{\mu\nu} =
 m^2 e^{-2\tilde A + 2H} \psi_{\mu \nu}
 \label{scho}
 \eeq
with
 \beq
 U(y)= e^{-2\tilde A + H/2} (e^{2\tilde A - H/2})^{\prime \prime}
 \eeq
As usual, in order to get a Schr\"{o}dinger-like equation, one must
change to conformal coordinates. However, from (\ref{scho}) we see
that the zero mode is normalizable
 \beq
 \psi_0^g \sim \,e^{2\tilde A - H/2}
 \eeq
and since $U \to 0 $ at $|y|\to\infty $, there is no gap between the
massless and the massive modes. This can be seen by writing  $U$ in
terms of $W$
 \beq
 U(y)= e^{H/2} (e^{-H/2})'' - e^{2H}\left[ 6 \left(\frac{d  \tilde{W}}{d\phi}\right)^2 - 4 \tilde{W}^2\right]
 \eeq
Clearly, shifting $W$ by a constant does not change the asymptotic
behavior of $U$ and the massive modes are therefore expected to
decouple \cite{Csaki:2000fc}. The zero mode, in proper length
coordinates, is shown in Fig. \ref{fig2}. It follows that the
graviton zero mode gets localized on the sub-wall located at the
edge of the wall close to the region of smaller $AdS_5$ curvature,
thus defining the Planck sub-wall. A similar localization of the
graviton zero mode on a rather different asymmetric double wall
system was found in \cite{Guerrero:2005aw}.

\subsection{Adding fermions}

Fermion confinement in the double walls
(\ref{gaugemetric},\ref{doublesol}) was studied in
\cite{Melfo:2006hh}. In these coordinates, 5-dimensional spinors
coupled to the scalar field by a Yukawa term of the form
 \beq
\lambda \bar\psi \psi \, \phi
 \eeq
give  confined chiral fermion modes on the wall
 \beq
 \psi^f_0 \sim \, e^{- 2 \tilde A(y) } e^{-\lambda \int \phi(y)  e^{H(y) } dy }
 \eeq
for sufficiently large $ \lambda$. It was found in
\cite{Melfo:2006hh} that in general thin walls require large Yukawa
couplings, which in the asymmetric case are bounded from below by
the largest cosmological constant, in order to confine fermions. For
the double-brane system (\ref{adw}), however, the Yukawa coupling
depends also on $\delta$ and we get
 \beq
 \lambda   >\frac{2\sqrt{|\Lambda_+|}}{3 \pi} \frac{s}{\sqrt{\delta(s - 1/2)}}
 \eeq
Therefore, the Yukawa coupling can be kept at reasonable values if
the thickness of the sub-branes $\gamma$ is decreased (recall that
$\gamma\rightarrow 0$ for $s\gg 1$) while the separation between
them (the brane thickness $\sim\delta/\alpha$) is increased.

The equation for the fermion zero modes was integrated numerically,
and results are given in Fig. \ref{fig2}. As argued above, the
fermion zero modes get located in the opposite sub-wall to the
Planck sub-wall, thus defining the weak brane.

\subsection{Newtonian potential}

 In order to
have an effective four dimensional gravity on the weak sub-wall, we 
  should
demonstrate that the massive graviton modes decouple.
Before presenting the results for the double wall system, it is instructive to consider the case $s=1$, a single wall without reflection symmetry, with different cosmological constants $\Lambda_+$ and $\Lambda_-$ on both sides. A similar case was studied in \cite{Castillo-Felisola:2004eg} in the limit  of a very small asymmetry. In the general case, the massive modes in the small $m/\sqrt{\Lambda_-}$ approximation  are given by
\beq
\psi_m = \kappa  \, \frac{m^{5/2}}{ \Lambda_-^{5/4} }
\left[  \frac{6 m^2} { \Lambda_-} +  \left(\sqrt{\frac{\Lambda_+}{\Lambda_-}} - 1 \right)\right ]^{-1}
\eeq
with $\kappa $  of order one for any value of the parameters. In the reflection-symmetric case $\Lambda_-=\Lambda_+$, this modes give the well-known contributions to the Newtonian potential between masses separated by a 4-D distance $r$ proportional to $1/r^3$. However, in  the asymmetric brane  the modes behave as $m^{5/2}$, and their  contributions are proportional to $1/r^7$, the equivalent of having 6 extra compact dimensions. We expect that the double asymmetric walls exhibit a similar behavior.

To calculate the contribution of the massless and massive modes to
the Newtonian potential, we switch to conformal coordinates
 \beq\label{toy-metric-conf}
ds^2= e^{2A(z)}\left(-dt^2 + dx^idx^i + dz^2\right)
 \eeq
and consider the $s\rightarrow \infty$ limit, i.e. the infinitely
thin sub-wall idealization of (\ref{adw}). This limit is not as
straightforward as (\ref{inf-double}) for the symmetric case
(\ref{doublesol}), due to the presence of the hypergeometric
function in the warp factor. However, it can be approximated by
 \beq\label{toy-conf}
 e^{A(z)}=\begin{cases}
        \left[e^{+\beta\delta/\alpha}+(\alpha+\beta)(z-z_+)\right]^{-1},& z_+ < z \cr
       \left[\cosh\beta\delta/\alpha + \beta z\right]^{-1},&z_-< z < z_+ \cr
        \left[e^{-\beta\delta/\alpha}-(\alpha-\beta)(z-z_-)\right]^{-1},& z < z_-\cr
        \end{cases}
 \eeq
with $z_\pm= \pm\beta^{-1}\sinh\beta\delta/\alpha$, $0<\beta<\alpha$
and $\delta>0$. For $\beta\rightarrow 0$ we obtain
(\ref{inf-double}),  the limit $s\rightarrow \infty$ of the
symmetric double wall (\ref{gaugemetric}, \ref{doublesol}), written
in conformal coordinates.

The Einstein tensor of (\ref{toy-metric-conf}, \ref{toy-conf}) is
given by
 \begin{eqnarray}
 \label{toy-einstein}
 G^a_{\,\,b} + \Lambda g^a_{\,\,b} &=& -\left(\partial^adt_b +
 \partial^b_{x^i}dx^i_b\right) \nonumber \\
 & &\left[ \sigma_+\,\delta(z-z_+) +
 \sigma_-\,\delta(z-z_-)\right]
 \end{eqnarray}
with $\Lambda$ given in each region by
  (\ref{asimcosmo}) and where the sub-brane tensions are
$\sigma_+=3\alpha e^{+\beta\delta/\alpha}$ and $\sigma_-=3\alpha
e^{-\beta\delta/\alpha}$. Notice that although this resembles the two
positive tension three-branes scenario of  
\cite{Lykken:1999nb}, in our case the two branes separate $AdS_5$
slices with different cosmological constants. Furthermore, the weak brane is not a so-called ``probe'' brane since its tension is not small (in fact, in proper length coordinates both sub-wall tensions are equal).

Now,   writing the transverse-traceless part of the metric
perturbations $h_{\mu\nu}$ as
 \beq
 h_{\mu\nu}(x,z)   = e^{i p\cdot x}  e^{A(z)/2 } \psi^g_{\mu\nu}(z)
 \eeq
with $h_{\mu z}=0$, one finds that the gravitational modes
$\psi^g_{\mu\nu}$ satisfy the Schr\"{o}dinger equation
 \beq
 \left(-\frac{d^2}{dz^2} + V_{QM}\right)\psi^g_{\mu\nu}(z)=
 m^2\psi^g_{\mu\nu}(z)
 \label{schoconf}
 \eeq
where
 \beq
 V_{QM}= \left( e^{3 A(z)/2}\right)'' e^{-3 A(z)/2}
 \eeq

For $m^2=0$ the solution is
 \beq
 \psi^g_0(z)=  N^g_0e^{3A(z)/2}
 \eeq
with
 \beq
  N^g_0= \left[ \frac{e^{ 2\beta\delta/\alpha}}{2(\alpha -\beta)} +
  \frac{e^{ -2\beta\delta/\alpha}}{2(\alpha + \beta)}   +
  \frac{1}{\beta}  \sinh(2\beta\delta/\alpha)  \right]^{-1/2}
 \eeq
The massive modes of (\ref{schoconf}) are given by 
\begin{widetext}
   \beq
 \psi^g_m(z)=N^g_m\begin{cases}
        (k_+^{-1}+ |z-z_+|)^{1/2}\left[ Y_2(m(k_+^{-1}+ |z-z_+|))+ C_+J_2(m(k_+^{-1}+|z-z_+|))\right],& z_+ < z \cr
         & \cr
        (k_0^{-1}+ z)^{1/2}\left[ A\,Y_2(m(k_0^{-1}+ z))+ B\,J_2(m(k_0^{-1}+z))\right],&z_-< z < z_+ \cr
         & \cr
        (k_-^{-1}+ |z-z_-|)^{1/2}\left[ Y_2(m(k_-^{-1}+ |z-z_-|))+ C_-J_2(m(k_-^{-1}+|z-z_-|))\right],& z < z_-\cr
        \end{cases}
 \eeq
 \end{widetext}
where $Y_2$ and $J_2$ are the Bessel functions of order two, 
$k_\pm=
(\alpha \pm \beta) \exp\{\mp\beta\delta/\alpha\}$, $k_0= \beta [\cosh(\beta\delta/\alpha)]^{-1}$ and
$C_+, C_-, A, B$ are constants determined by the matching conditions.
 
The contribution of the massive modes to the Newtonian potential can
be now calculated by expanding around masses smaller than the
smaller scale of the system, which is of order $\beta
e^{-\beta\delta/\alpha}$ (notice that the symmetric case $\beta=0$ has to be treated separately).  The normalized wave functions for the
massive modes in the weak and Planck branes are found to be
 \beq
  \psi^g_m(z_+)= \frac{\kappa_+}{4}
  \left[{\frac{\beta}{5}\left(1-\frac{\beta^2}{\alpha^2}\right)}\right]^{1/2}
  \left(\frac{m}{\alpha \,e^{-\beta\delta/\alpha}}\right)^{5/2}
   \eeq
   \beq
    \psi^g_m(z_-)= \frac{\kappa_-}{4}
  \left[{\frac{\beta}{5}\left(1-\frac{\beta^2}{\alpha^2}\right)}\right]^{1/2}
  \left(\frac{m}{\alpha \,e^{-\beta\delta/\alpha}}\right)^{5/2}\,
    e^{3 \beta\delta/2 \alpha}
\eeq to leading order in $e^{-\beta\delta/\alpha}$, were $\kappa_\pm$
are functions of $\alpha$ and $\beta$ of order unity. Here, in order
to obtain the correct normalization constant for the wave function
of the massive modes, we have used two regulator branes with
positions at $\pm z_r$ with $z_r > |z_{\pm}|$ and taking the limit
$z_r\rightarrow \infty$ at the end of the calculation, extending to
our two brane system, as close as possible, the single brane
treatment of \cite{Callin:2004py}.
We can see that the modes behave as $m^{5/2}$, as in the single  asymmetric brane, and therefore for $r<r_c$ the Newtonian potential behaves as in a scenario with six extra dimensions. 

The Newtonian potential   in the weak brane is
 \beq
V_+(r) = \frac{G^{\rm W} m_1 m_2}{r} \left[1 + \kappa_+^2
\left(\frac{r_c}{r}\right)^6 \right]
 \eeq
and in  the Planck brane
 \beq
 V_-(r)=  \frac{G^{\rm P} m_1 m_2}{r}  \left[1 +  \kappa_-^2
 \left(\frac{r_c}{r}\right)^6   \right]
 \eeq
to leading order in $e^{- \beta\delta/\alpha}$, where $r_c=
\alpha^{-1}\exp{(5\beta\delta/3\alpha)}$ and the four-dimensional
gravitational constants are given by
 \beq G^{\rm W}_N =
G_5  (N_0^g)^2 e^{ -3\beta\delta/\alpha},\quad G^{\rm P}_N = G_5
(N_0^g)^2 e^{ 3\beta\delta/\alpha}\,
 \eeq
The ratio of the Planck masses on both branes is then
 \beq
 M_{Pl}^{\rm Weak}/M_{Pl}^{\rm Planck}= e^{ 3 \beta\delta/\alpha}
 \eeq
and we can now refine our result (\ref{supress}). For $M_{Pl}^{\rm
Planck} $ of order TeV, we would get a Planck mass in the weak brane
of order $10^{19}$ GeV by setting
 \beq \frac{\beta}{\alpha}\,\delta \simeq 12
 \eeq

\section{Summary and outlook}

We have shown that the large hierarchy between the Planck and the
weak scales can be attributed to the tendency of gravitons and
fermions to localize at different edges of an asymmetric double
domain wall. The embedding in a five dimensional Anti-de Sitter
geometry which is not reflection symmetric determines which
sub-brane is the weak brane and which one is the Planck brane.
Fermions of one chirality  are localized by their Yukawa
interactions with the scalar field, but the warped metric forces
them to live in a different wall than the gravitons, without need
for additional assumptions. By calculating the massive mode
contributions in this system, we have shown that Newtonian gravity
is recuperated in the sub-wall where the matter is located.

The double-wall systems are straightforward generalizations of
single, reflection symmetric BPS walls, and while here we have
considered the case of the simplest one, the confinement of graviton
and fermion zero modes on different sub-walls of the system is
presumably generic to other double solutions without reflection symmetry. 
The generalization
consists simply in taking a power of the fake superpotential, and
then adding a constant to it. Since the double asymmetric walls are
also BPS and therefore have a topological charge (which is in fact
the same as the original wall), stability is guaranteed, and no
additional fields or stabilization mechanism are required.

While we believe this to be an interesting effect, many questions
would have to be answered before attempting to use it as a solution
to the hierarchy problem. For example, a mechanism for confinement
of the gauge fields would be required, in particular, one that makes
use of the fermionic fields to achieve localization such as in \cite{Dvali:1996xe}   would be
well-suited, since then gauge fields will be confined to the matter
wall. Scalar fields are also a problem, since the asymmetry drives
them to the graviton's wall, their modes being proportional to the
warp factor. An adequate coupling with the wall's scalar field could
help.  We hope to address this issues in a
future publication.

\section*{Acknowledgments} 

We wish to thank Borut Bajc, Goran Senjanovi\'c
and Rafael Torrealba for enlightening discussions, and W. Bietenholz for 
useful comments on the manuscript. This work was
supported by CDCHT-UCLA project 006-CT-2005, by CDCHT-ULA project
No. C-1267-04-05-A and  by FONACIT projects S1-2000000820 and
F-2002000426. A.M. thanks ICTP for hospitality during the completion
of this work.


\end{document}